\documentclass[prd,twocolumn,floatfix,preprintnumbers]{revtex4}

\usepackage{graphicx}
\input{epsf}
\usepackage{epsf}
\usepackage{graphicx,epsfig}
\usepackage{bm}
\usepackage{latexsym,amssymb,amsmath,float}

\newcommand {\ga} {\ {\raise-.5ex\hbox{$\buildrel>\over\sim$}}\ }
\newcommand {\la} {\ {\raise-.5ex\hbox{$\buildrel<\over\sim$}}\ }

\def\be{\begin{equation}}
\def\ee{\end{equation}}
\def\ba{\begin{eqnarray}}
\def\ea{\end{eqnarray}}

\begin{document}

\title{Big Bang nucleosynthesis with a stiff fluid}
\author{Sourish Dutta and Robert J. Scherrer}
\affiliation{Department of Physics and Astronomy, Vanderbilt University,
Nashville, TN  ~~37235}

\begin{abstract}
Models that lead to a cosmological stiff fluid component,
with a density $\rho_S$ that scales as $a^{-6}$, where $a$ is
the scale factor, have been proposed recently in a variety of contexts.
We calculate numerically the effect of such a stiff fluid
on the primordial element abundances.
Because the stiff fluid energy density decreases with the scale factor
more rapidly than radiation,
it produces a relatively larger change in the primordial helium-4 abundance
than in the other element abundances, relative to the changes produced by
an additional radiation component.  We show that the helium-4 abundance
varies linearly with the density of the stiff fluid at a fixed fiducial temperature.
Taking $\rho_{S10}$ and $\rho_{R10}$ to be the stiff fluid energy density and the
standard density in relativistic particles, respectively, at $T = 10$ MeV,
we find that the change in the primordial helium abundance is well-fit by
$\Delta Y_p = 0.00024(\rho_{S10}/\rho_{R10})$.  The
changes in the helium-4 abundance produced by additional radiation or by a stiff fluid
are identical when these two components have equal density at a ``pivot temperature", $T_*$,
where we find $T_* = 0.55$ MeV.  Current estimates of the primordial $^4$He abundance
give the constraint on a stiff fluid energy density of
$\rho_{S10}/\rho_{R10} < 30$.

\end{abstract}

\maketitle

\section{Introduction}

In recent years, a host of cosmological observations have provided an increasingly precise picture
of the constituents of the universe.  The baryon density has long been known to
provide roughly 5\% of
the critical density; earlier estimates from Big Bang nucleosynthesis
\cite{redux} have been confirmed
by CMB observations from WMAP \cite{WMAP}.
Cosmological data from a wide range of sources including type Ia supernovae \cite{hicken},
the cosmic microwave background \cite{WMAP}, baryon acoustic oscillations \cite{percival},
cluster gas fractions \cite{Samushia2007,Ettori} and gamma ray bursts \cite{Wang,Samushia2009} seem
to indicate that about 72\% of the energy density of the Universe
is in the form of an exotic, negative-pressure component called dark energy. In addition,
a slew of evidence from modern sources, including weak \cite{weak} and strong \cite{strong} lensing,
the bullet cluster \cite{bullet}, Large Scale Structure \cite{lss}, as well as supernovae and the CMB,
have confirmed  earlier indications from rotation curves \cite{old1,old2,old3} that another 23\%
of the energy density of the Universe is in the form of weakly interacting matter called dark
matter.  (See Ref. \cite{Copeland} for a review of dark energy and \cite{feng}
for a review of dark matter). 

While the existence of all of these components is reasonably well-established,
the existence of other exotic fluids
is not ruled out by the current data. For example, several models predict the existence of ``dark
radiation''  either early or late in the history of the Universe (see e.g. \cite{darkrad} for an example
of such a model and references therein for others). Another exotic fluid which arises in
various models is a ``stiff fluid'', i.e.,
a fluid with an equation of state parameter $w_S \equiv p_S/\rho_S =1$.
This is the largest value of $w$ consistent with causality, since the speed of sound of this fluid equals the speed of light. Such models were apparently first studied by Zeldovich \cite{zeldovich}. The Friedman equation for such a fluid implies that its energy density $\rho_S$ varies with the scale factor $a$ as:
\be
\rho_S\propto a^{-6}.
\ee

In recent years, a variety of models have been proposed that produce a stiff cosmological fluid:

\paragraph{Kination:} A kination field is a scalar field whose energy density is dominated by kinetic
energy. A period of kination can follow a period of inflation, and was
first studied in the context of electroweak baryogenesis \cite{kination,kination1}.
Its impact on reheating as well as on the freeze-out of dark matter particles has
been studied in Refs. \cite{kamionkowski,salati,pallis1,gomez,pallis2}. Kination fields have also been applied
to unify inflation and dark energy by using the same scalar field for both
inflation and quintessence. In these models inflation ends with a period of
kination as the inflaton receives a ``kick'', and the same field later on plays the role of quintessence \cite{everett}. 

\paragraph{Interacting dark matter:} In models with a warm self-interacting dark matter component, the elastic self-interactions between the (scalar boson or fermionic) dark matter particles can be characterized by the exchange of vector mesons via minimal coupling. For these models the self-interaction energy can be shown to behave like a stiff fluid \cite{selfint}.

\paragraph{Ho\v{r}ava-Lifshitz cosmologies:} Stiff fluids also occur
in certain cosmological models based on the recently proposed Ho\v{r}ava-Lifshitz gravity,
a power-counting renormalizable and ultraviolet-complete field theoretic quantum gravity
model based on "anisotropic scaling" of the space and time dimensions \cite{hor2,hor1,hor3,hor4}.
In the original formulation of this theory a ``detailed balance'' condition was
imposed as a convenient simplification \cite{hor3}. The  validity and usefulness of
the detailed balance condition have subsequently been discussed extensively
(see e.g. \cite{calcagni,kiritsiskofinas}) as well as the consequences of relaxing it \cite{relax1,relax2,relax3,relax4}. Stiff fluids arise in models in which the detailed balance condition has been relaxed. 
Cosmological models based on  Ho\v{r}ava-Lifshitz gravity have been studied extensively (see e.g. \cite{HLcosmo1,HLcosmo2,HLcosmo3,HLcosmo4,HLcosmo5,HLcosmo6}) and observational constraints on such models, including the stiff-fluid cases, were considered in \cite{duttasaridakis1,duttasaridakis2,duttasaridakis3}. However, it must also be noted that the theoretical foundations of Ho\v{r}ava-Lifshitz gravity are still under debate (see e.g. \cite{crit1,crit2,crit3,crit4,crit5,Koyama:2009hc,Papazoglou:2009fj,Kimpton:2010xi,Bellorin:2010je}).

\paragraph{Non-singular cosmological models:} Stiff fluids have been also found to show up as exact non-singular solutions in inhomogeneous cosmological models \cite{stiffsol1,stiffsol2,stiffsol3,stiffsol4}.

Given the recent flurry of interest in such models, it is clearly useful to derive precise limits
on the density of a stiff fluid in the early universe.
Because the density of a stiff fluid decays more rapidly than either radiation or matter, the effect
of the stiff fluid on the expansion rate will be the largest at early times.  Thus, the strongest
limits on the density of such a fluid come from Big Bang nucleosynthesis (BBN), which remains
the earliest cosmological process whose evolution can be determined with high precision.
(Note that some attempts have been made to constrain the expansion rate of the universe
prior to BBN from the relic
dark matter abundance \cite{kamionkowski,pallis1,gomez,donato,drees,arbey1,arbey2}.  While the freeze-out of the dark matter does occur at
a much earlier time than BBN, the exact model for the dark matter is less precisely determined).

Previous discussions of stiff fluids have usually quoted BBN limits on the expansion rate
at a fixed temperature (typically $T \sim 1$ MeV) and used these limits to constrain the density
of the stiff fluid.  Here, we numerically evolve the element abundances in the presence of the stiff fluid to
derive the exact dependence of the element abundances on the stiff fluid density.
The stiff fluid investigated here resembles a special case of the models examined by
Masso and Rota \cite{masso} who investigated the effect of an arbitary additional density
of the form $\lambda (T/0.1~{\rm MeV})^\gamma$, although their $\gamma = 6$ case
is not identical to the model considered here,
since $T$ does not scale as $1/a$ through the era of
$e^+e^-$ annihilation.  Here, we examine the stiff fluid case in more detail, and we exploit
the fact that WMAP now provides an independent determination of the baryon-photon
ratio \cite{WMAP},
effectively eliminating a degree of freedom from BBN and allowing for better constraints on the
stiff fluid.
(See also related work on a somewhat different variant model in Ref. \cite{carroll}).
We present our
calculations in the next section, and discuss our limits in Sec. III.

\section{Effect of a stiff fluid on BBN}

BBN has long been used to constrain additional energy density in the early universe (for recent
reviews, see Refs. \cite{serpreview,steigreview,iocreview}).  The expansion rate $H$ is given by
\begin{equation}
H^2 = \frac{8 \pi G}{3} \rho,
\end{equation}
where $\rho$ is the total density, so any additional contribution to $\rho$ increases the expansion rate
and changes
the resulting element abundances.

At high temperatures ($T \ga 1$ MeV) the rates for the weak
interactions which govern the interconversion of neutrons and protons,
\begin{eqnarray}
n+\nu _{e} &\leftrightarrow &p+e^{-},  \notag \\
n+e^{+} &\leftrightarrow &p+\bar{\nu}_{e},  \notag \\
n &\leftrightarrow &p+e^{-}+\bar{\nu}_{e},
\end{eqnarray}%
are
larger than the expansion rate, $H$, and the neutron-to-proton ratio ($n/p$)
tracks its equilibrium value. As the universe expands and cools, the
weak interaction rates drop below the expansion rate,
and $n/p$ freezes out at $T \sim$ 1 MeV.  Between $T \sim
1$ MeV and $T \sim 0.1$ MeV,
the neutrons undergo free decay, and then at $T \sim 0.1$ MeV, nucleosynthesis
proceeds to fuse the remaining neutrons with the protons to produce heavier elements,
primarily $^4$He, but also trace amounts of $^2$H, $^3$He, and $^7$Li.
The final $^4$He abundance is most sensitive to the expansion rate near $T \sim
1$ MeV,
when the neutron/proton ratio freezes out, while the other element abundances are more
sensitive to the expansion rate near $T \sim 0.1$ MeV, when fusion into heavier elements
begins.

The dependence of the different element abundances on the expansion rate at
various temperatures was explored quantitatively by Bambi et al. \cite{bambi},
who derived ``response functions" that show the change in the abundance of each nuclide
as a function of a change in the expansion rate at a given temperature.
[Note that we use these response functions only to gain insight into the effects
of the stiff fluid on the element abundances; our actual calculation of the element abundances
utilizes a full numerical integration of the BBN equations, as discussed below].
Consider an additional source of energy density which changes the expansion
rate, $H(T)$, by an amount $\Delta H(T)$.  Then Bambi et al. argued that,
for a given value of the baryon-photon ratio $\eta$,
the change in a given nuclide abundance, $\Delta X_i$, is given by
\begin{equation}
\label{response}
\Delta X_i = 2 \int \varrho_i(T) \frac{\Delta H(T)}{H(T)} \frac{dT}{T},
\end{equation}
where $\varrho_i(T)$ is the response function for a given nuclide,
derived numerically for several elements of interest in Ref. \cite{bambi}.
As expected, the response functions for deuterium and
$^7$Li are strongly peaked near $T \sim 0.1$ MeV.  In contrast,
the response function for $^4$He is broadly distributed
between 1 MeV and 0.1 MeV, with
two peaks of roughly equal magnitude at these two temperatures
\cite{bambi}.  The
first corresponds to $n-p$ freeze-out, and the second to the onset of
fusion.  The latter affects the $^4$He abundance primarily through the influence
of free-neutron decay; the earlier fusion begins, the more undecayed neutrons
remain to form $^4$He.

Now consider a model, such as the one examined here,
with some additional source of energy density,
$\Delta \rho(T)$.
As long as $\Delta \rho(T) \ll \rho(T)$, where $\rho(T)$
is the standard energy density, we can write equation (\ref{response})
as
\begin{equation}
\label{response2}
\Delta X_i = \int \varrho_i(T) \frac{\Delta \rho(T)}{\rho(T)} \frac{dT}{T}.
\end{equation}
Because of the sensitivity of the element abundances to the density
at different epochs, the relative changes in the different element
abundances will be different for different functional forms of $\rho(T)$.
However, consider an arbitrary functional dependence of the form
$\rho(T) = \rho(T_0) f(T/T_0)$, where $\rho(T_0)$ is the density at some
fixed fiducial temperature $T_0$, while $f(T/T_0)$ is
an arbitrary function subject only to the constraint $f(1) = 1$.  Thus, $f(T/T_0)$ parametrizes
the dependence
of $\rho$ on $T$, while $\rho(T_0)$ fixes the overall amplitude of the density. 
The important point is that as long as Eq. (\ref{response2}) is a good
approximation, and once $f(T/T_0)$ is fixed, the abundance of
each nuclide will vary linearly with $\rho(T_0)$, regardless of the functional
form of $f(T/T_0)$.

This linear dependence is seen in the case
of additional relativistic energy density
scaling
as $\Delta \rho_R \propto a^{-4}$.  Parametrizing this energy density
in terms of the number of additional two-component
neutrinos, $\Delta N_\nu$, the change in the primordial $^4$He mass
fraction, $\Delta Y_p$, is well approximated by \cite{kneller}
\begin{equation}
\label{nu}
\Delta Y_p = 0.013 ~\Delta N_{\nu}.
\end{equation}

Given the different sensitivities of the element abundances to the
expansion rate at different temperatures, it is clear that 
the change in the element abundances produced by a stiff fluid
will differ from that produced by additional relativistic energy density.
However, if we confine our attention to a single element (such as $^4$He), then
we expect the overall abundance to scale linearly with the value of the
stiff energy density at a fixed fiducial temperature.

We model our stiff fluid as a component with energy density $\rho_S$, given by
\begin{equation}
\label{rhos}
\rho_S = \rho_{S10}(a/a_{10})^{-6}.
\end{equation}
where $\rho_{S10}$ and $a_{10}$ are the stiff fluid density and scale factor, respectively,
at $T = 10$ MeV (well before $e^+e^-$ annihilation).  
We use the Kawano \cite{kawano} version
of the Wagoner \cite{wagoner1,wagoner2} nucleosynthesis code to derive the
element abundances as a function of $\rho_{S10}$ and
of the baryon-photon ratio, $\eta$.

In Fig. 1 we compare the change in the element abundances produced by a stiff fluid
with that from one extra neutrino species, for $\eta$ in the range $5 - 7 \times 10^{-10}$.
The blue (solid) curve gives the standard BBN model with no additional energy density.
The black (dashed) curve
denotes the element abundances due to one additional two-component neutrino, while the
red (dotted) curve gives the abundances due to a stiff fluid for which $\rho_{S10}$
is chosen to
produce the same effect on $Y_p$ as one extra neutrino
(as can be seen in the top panel).

As expected, when the stiff fluid density is adjusted to give the same effect on the $^4$He abundance
as an extra neutrino, the stiff fluid produces a much smaller effect on the deuterium and $^7$Li abundances.
This is because, as we have noted, the latter two element abundances are sensitive to the expansion rate at a much lower temperature
than is the $^4$He abundance, and the stiff fluid density decays with expansion rate much more rapidly than
does the contribution of an extra neutrino.

The one minor surprise is that the stiff fluid increases
the $^7$Li abundance, while an additional neutrino decreases it.  This can be understood
in terms of the response function in Ref. \cite{bambi}.  The $^7$Li response function has a sharp
trough just below $T = 0.1$ MeV, but it also has a shallow positive plateau for $T > 0.1$ MeV.
The physical reason for this behavior comes from the
fact that the dominant reaction for $^7$Li production is $^4$He + $^3$He
$\rightarrow ^7$Be + $\gamma$ \cite{serpreview}.  An increase in the expansion rate at high temperatures ($T > 0.1$ MeV) gives a larger
neutron abundance prior to nuclear fusion, enhancing the abundance of both $^4$He and $^3$He to make more
$^7$Be.  An increase in the expansion rate just below $T = 0.1$ MeV, when nuclear fusion is occurring,
gives less time for the fusion reactions to build into heavier elements; thus, the decrease in the $^7$Li
abundance in this case is
accompanied by increases in the abundances of deuterium and helium-3.
The stiff fluid samples the positive plateau at high $T$ more strongly than the
trough at low $T$, while the reverse is true for
an extra neutrino.  

Once $\eta$ is fixed by the CMB, the best constraint on the stiff fluid can be obtained
from the $^4$He abundance.  WMAP7 \cite{WMAP} gives $\eta = 6.2 \times 10^{-10}$.  For this
value of $\eta$, we plot,
in Fig. 2, the change in the primordial $^4$He abundance, $\Delta Y_p$
as a function of $\rho_{S10}/\rho_{R10}$,
where $\rho_{S10}$ is the density of the stiff fluid at $T = 10$ MeV (as in equation \ref{rhos}),
while $\rho_{R10}$ is the standard energy density in relativistic particles
at $T = 10$ MeV.
As expected, we find that $\Delta Y_p$ is well-fit by a linear dependence on
$\rho_{S10}/\rho_{R10}$, namely
\begin{equation}
\label{rhoS}
\Delta Y_p = 0.00024 (\rho_{S10}/\rho_{R10}).
\end{equation}
We find further that Eq. (\ref{rhoS}) is an excellent approximation
when $\eta$ lies in the range
between $5 \times 10^{-10}$ and $7 \times 10^{-10}$.  Of course, higher-order
corrections produce a slight deviation from exactly linear behavior, as is evident
in Fig. 2 (see, e.g., Appendix F of Ref. \cite{serpreview} for a discussion of such
corrections for $\Delta N_\nu$).

\begin{figure}
\begin{center}
	\epsfig{file=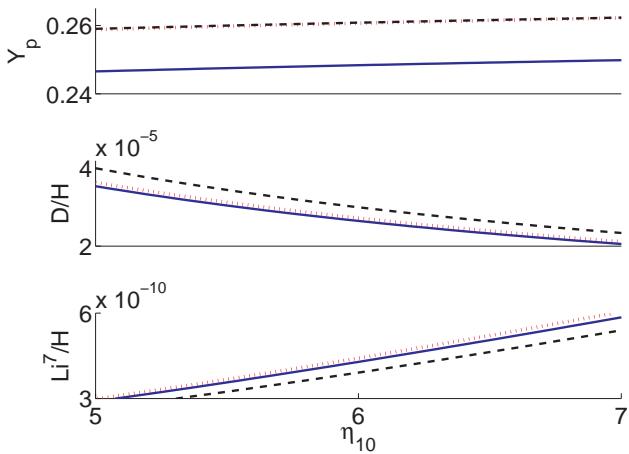,height=60mm}
	\caption
	{\label{comparison}{Changes in primordial element
	abundances caused by a stiff fluid and by an extra
	neutrino, as a function of $\eta_{10}\equiv \eta \times 10^{10}$,
	where $\eta$ is the baryon-photon ratio. The black
	(dashed) curve denotes the element abundances due to
	one extra two-component neutrino, the red (dotted) curve
	denotes the abundances for a stiff fluid, and the blue
	(solid) curve denotes the unmodified standard case
	of 3 neutrinos and no stiff fluid.
	The density of the stiff fluid has been adjusted
	to produce the same effect on $Y_p$ as one extra
	neutrino, as can be seen from the top panel. }}
\end{center}
\end{figure}

\begin{figure}
\begin{center}
	\epsfig{file=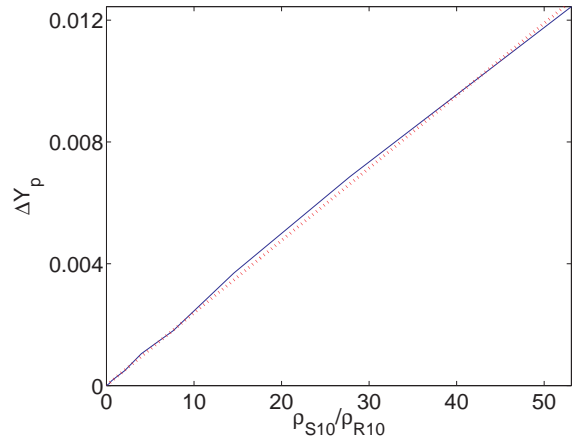,height=60mm}
	\caption
	{\label{prefac_Yp}{The change in the primordial $^4$He abundance,
	$\Delta Y_p$, produced by a stiff fluid with density
	$\rho_S = \rho_{S10}(a/a_{10})^{-6}$, where $\rho_{S10}$ is
	the stiff fluid density at a temperature of 10 MeV, and $a_{10}$
	is the scale factor at this temperature.  The horizontal axis shows
	$\rho_{S10}/\rho_{R10}$, where
	$\rho_{R10}$ is the total energy density in relativistic particles
	(i.e., in the standard cosmological model without a stiff fluid)
	at 10 MeV}.  Blue (solid) curve is numerical result, and red (dotted)
	curve is the fit from Eq. (\ref{rhoS}).  These results are for a baryon-photon ratio
	of $\eta = 6.2 \times 10^{-10}$.}
\end{center}
\end{figure}

\section{Discussion}

The strongest constraints on a stiff fluid clearly come from the primordial $^4$He abundance.
This abundance remains at present somewhat uncertain (for recent discussions, see, e.g.,
Refs. \cite{izotov,aver,steigman}).  Recent analyses by Izotov and Thuan \cite{izotov}
and by Aver, Olive, and Skillman \cite{aver}
are both consistent with
a central value of $Y_p = 0.256$.  Using this limit with the WMAP7 value of
$\eta = 6.2 \times 10^{-10}$, we obtain the bound
\begin{equation}
\rho_{S10}/\rho_{R10} < 30.
\end{equation}

This result, however, should not be considered the main result of our paper, as the estimates of
$Y_p$ are likely to improve with time.  Rather, our main conclusion is embodied
in Eq. (\ref{rhoS}), a
result that can be used to provide an upper bound on the stiff fluid density for any
estimate of the
primordial $^4$He abundance.

We can exploit the fact that $\Delta Y_p$ depends linearly on both $\Delta N_\nu$ and
$\rho_{S10}/\rho_{R10}$ to derive a ``pivot temperature," $T_*$, at which equal contributions
from relativistic energy density or from a stiff fluid will produce equal changes in
$^4$He.  In other words, suppose that we have a particular
value of $\Delta Y_p$.  This will correspond to a particular value of $\Delta
N_\nu$ in Eq. (\ref{nu}), or to a particular value of $\rho_{S10}/\rho_{R10}$ in Eq. (\ref{rhoS}).  In this case,
the additional energy densities in radiation or in the stiff fluid will be equal at a single
temperature $T_*$:
\begin{equation}
\Delta \rho_R(T_*) = \rho_S (T_*).
\end{equation}
By comparing Eqs. (\ref{nu}) and (\ref{rhoS}), we find that
\begin{equation}
T_* = 0.55~ {\rm MeV}.
\end{equation}
We emphasize that this is a purely heuristic result. It is not based
on the assumption that the helium abundance depends only on the expansion rate at a single
temperature; we have seen that it does not. Indeed, it is not 
surprising that $T_*$
lies in between the two peaks in the $^4$He response function.  Note further that
$T_*$ corresponds, strictly speaking, to the neutrino temperature, rather than the photon temperature.
The neutrino temperature scales
exactly as $1/a$, while the photon temperature, even near 0.55 MeV, has experienced
a small increase due to $e^+e^-$ annihilation.  Since most bounds from BBN
on additional
energy
density are expressed in terms of limits on additional radiation density (or equivalently,
additional neutrinos)
from the primordial helium abundance, our result for the pivot temperature can be used
to convert these bounds into limits on an additional stiff fluid component.

\section{Acknowledgments}
We thank E. Linder, G. Steigman and and R. Stiele for helpful comments on the manuscript.
We thank E. Saridakis for useful discussions.
R.J.S. was supported in part by the Department of Energy
(DE-FG05-85ER40226).

\end{document}